\documentclass{article}
\usepackage{amsmath,amsfonts,amssymb}
\newcommand{\complex}{\kern.1em{\raise.47ex\hbox{
            $\scriptscriptstyle |$}}\kern-.40em{\rm C}}

\def\mathbb#1{{\bf #1}}

\newcommand{\Hm}{{\bf H}}

\title{Magnetic Fingerprints of Fractal Spectra and Duality of Hofstadter Models}%
\begin{document}
\input epsf
\author{ O. Gat and J.~E.~Avron\footnote{avron@physics.technion.ac.il, omri@physics.technion.ac.il} }
\maketitle
\centerline{Department of Physics, Technion, 32000, Haifa, Israel}
\begin{abstract}
We study the de-Haas van Alphen oscillations in the magnetization of the Hofstadter model. Near a split band the magnetization is a rapidly oscillating function of the Fermi energy with lip shaped envelopes. For generic magnetic fields this structure appears on all scales and provides a thermodynamic fingerprint of the fractal properties of the model. The analysis applies equally well to the two dual interpretations of the Hofstadter model and the nature of the duality transformation is elucidated.

\end{abstract}



%



The  Hofstadter model~\cite{ref:Hofstadter} describes non-interacting (spinless) electrons moving in the plane under the combined action of magnetic field and a periodic potential. 
The model is a paradigm for quantum systems with singular continuous spectra~\cite{ref:azbel-wilkinson,ref:last}; Chern numbers in the integer quantum Hall effect~\cite{ref:tknn}; Fractal quantum phase diagrams~\cite{ref:osadchy}; Quantum integrable models \cite{ref:wiegman-in} and Non-commutative geometry~\cite{ref:connes}. It has been realized experimentally in~\cite{ref:albrecht} and measurements of the Hall conductance phase diagram are in good agreement with theoretical predictions of \cite{ref:tknn}. 

Previous studies of the thermodynamics of the Hofstadter model have focused on minima of the ground state energy~\cite{ref:weigman}. Our aim is to examine the possibility that the de Haas-van Alphen oscillations, which are a basic tool in the study of the geometry of Fermi surfaces of metals~\cite{ref:aschroft-mermin}, also provide a tool for studying its non-commutative analog. {We show that when the magnetic flux per unit cell is close to rational multiple of the flux quantum,  the magnetization oscillates as a function of the chemical potential in the vicinity of the bands of the rational flux, as in the de Hass-van Alphen effect. As the deviation from rational flux becomes smaller, the frequency of the oscillations increases, so that in the limit of the deviation tending to zero the oscillations fill an area bounded by a limiting envelope. The envelopes have the following universal features: They are smooth functions of the chemical potential except for logarithmic pinching at a single point, associated with a logarithmic divergence of the density of states, and they are pinched linearly at the edges of the bands. These properties give the envelopes suggestive lip-like shapes. 

It follows from these results that at magnetic flux ratios which are irrational numbers well approximated by infinitely many rationals numbers, magnetic oscillations bounded by lip-shaped envelopes occur on all energy scales}. In this sense, the magnetic oscillations provide a fingerprint of the fractal spectra of the model.

A second issue of the Hofstadter model that we discuss is duality.  The Hofstadter model approximates the Schr\"odinger equation in two dual limits: When the magnetic field is a weak perturbation of a Bloch  band and remarkably also in the dual limit when the magnetic field is so strong that the periodic potential is a perturbation of the lowest Landau level. As we shall see, the thermodynamic properties of the dual limits are related by an interesting duality transformation of the thermodynamic potential, Eq.~(\ref{eq:omega-tb-ll}) below. We examine the consequences of the duality for the magnetization and the Hall conductance in some detail.

The spectral analysis of the Hofstadter model reduces to studying a family of one dimensional problems parameterized by a conserved quasi-momentum $\xi$. For the sake of concreteness and simplicity we shall consider the case where the periodic potential has square symmetry. In this case the corresponding Hamiltonian is:
\begin{equation}\label{eq:hofs-matrix}
\big(H(\varphi,\xi)\psi\big)(j)=\psi(j-1)+\psi(j+1)+2\cos (2\pi\varphi\,j+\xi)\,\psi(j), \end{equation} 
{It is known that $H$ describes the splitting of a Bloch band by a weak transverse magnetic field, and also the splitting of a Landau level by a weak periodic potential. $\varphi$ and $\varphi^{-1}$ are the values of flux through a unit cell, measured in units of quantum flux $\Phi_0=hc/e$, in the split Bloch and split Landau cases, respectively}. Note that $\varphi$ is a dimensionless quantity. 

Although  $H(\varphi,\xi)$ depends analytically on $\varphi$ its spectral properties are sensitive to the flux: When $\varphi$ is a rational number the spectrum is a finite collection of energy bands, while for (almost all) irrational fluxes it is a Cantor set of zero Lebesgue measure \cite{ref:last}.

\begin{figure}[h]
\epsfysize=2.5in\hskip -.4 in\hbox{\epsfbox{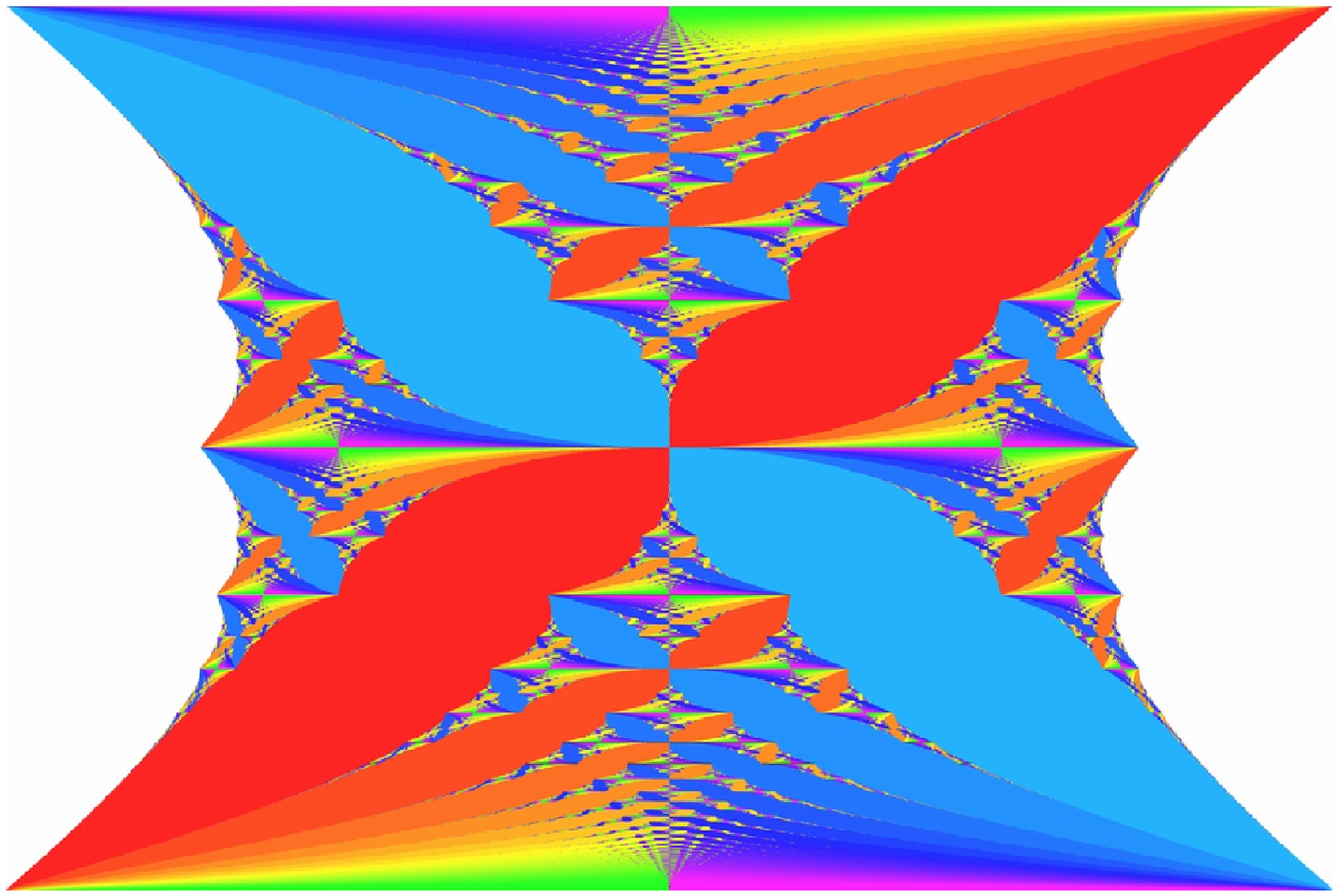} 
\hskip 0.in\epsfysize=2.5in \epsfbox{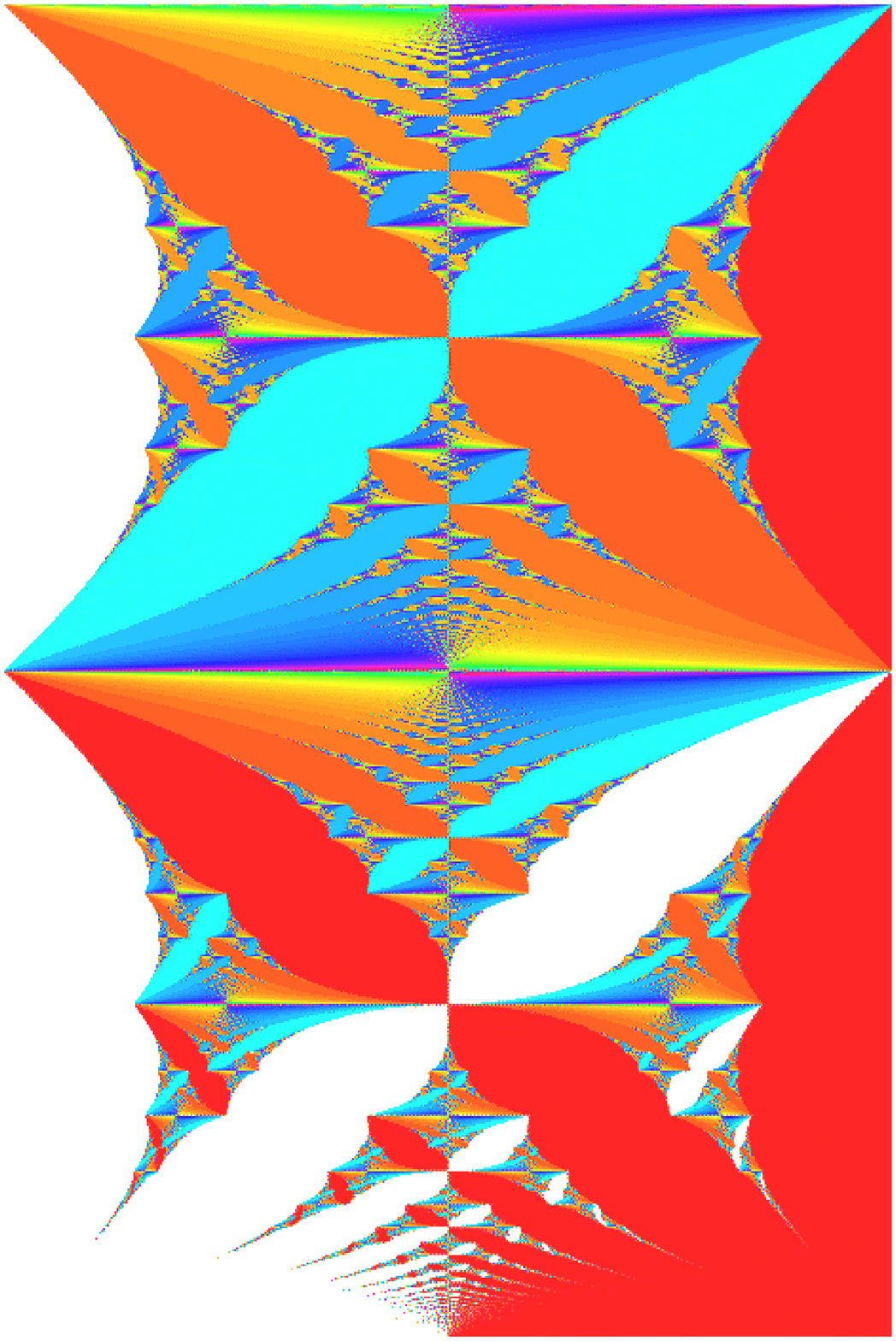}}
\bigskip 
\caption{The quantum phase diagrams of a split Bloch band (left) and split Landau level (right). The horizontal axis is the chemical potential and the vertical axis is $0\le\varphi\le1$ on the left and $0\le\varphi^{-1}\le2$ on the right. The colors represent the quantized Hall conductances. The scheme is such that cold colors correspond to negative integers and warm colors positive integers. Zero is white.  The red on the right hand side of the right diagram corresponds to unit Hall conductance of a full Landau band. } \label{fig:ll}
\end{figure} 
Let $\Omega_{b}(T,\mu,\varphi)$ be the thermodynamic potential per unit cell of a split Bloch band.  By gauge invariance, time-reversal and electron-hole symmetry, $\Omega_b$ satisfies 
\begin{eqnarray}\label{eq:symmetry-omega}
\Omega_b(T,\mu,\varphi)&=&\Omega_b(T,\mu,-\varphi)
\nonumber \\
&=&\Omega_b(T,\mu,\varphi+1)=\mu+\Omega_b(T,-\mu,\varphi)
\end{eqnarray}
It follows that the thermodynamic properties for all $\varphi$ and $\mu$ are determined by those for  $0\le \varphi\le1/2,\ \mu< 0$. $\mu=0$ corresponds to the half-filling point of electron-hole symmetry. {The phase diagram of the split Bloch band is displayed on the left hand side of Fig.~\ref{fig:ll}. The phases are labelled by the values of the Hall conductance, whose relation to $\Omega$ is given in Eq.~(\ref{eq:rh-m-s}) below. The phase diagram displays the symmetry properties shown in Eq.~(\ref{eq:symmetry-omega}). The phase diagram is periodic in $\varphi$, see Eq.~(\ref{eq:symmetry-omega}); a single period is given in Fig.~\ref{fig:ll}.

It is straightforward to arrive from the the thermodynamics of the split Bloch band to that of the split Landau level by observing that the energy spectra of the two models are the same. The difference is that the number of states per unit area in a split Bloch band is a constant independent of the magnetic field, whereas the number of states per unit area of a Landau level is proportional to the magnetic field. Therefore}
the thermodynamic potentials of a split Landau level, $\Omega_l$, and split Bloch band are related by 
\begin{equation}\label{eq:omega-tb-ll}
\Omega_l(T,\mu,\varphi)=\varphi\,\Omega_b(T,\mu,\varphi^{-1})
\end{equation}
This is a duality transformation: It is symmetric under the interchange \hbox{$b\longleftrightarrow l$}. It implies that the thermodynamics of the split Bloch band determine the thermodynamics of a split Landau level and vice versa. {It follows from Eq.~(\ref{eq:omega-tb-ll}) that $\Omega_l$ is not periodic, although $\Omega_b$ is, simply because of the factor $\varphi$ on the right. A part of the phase diagram of the split Landau level is shown on the right hand side of Fig.~\ref{fig:ll}, and it is evidently aperiodic.}

To derive the relation between the magnetization and Hall conductances of the two models recall that the filling fraction $\rho$, the magnetization per unit area, $m$, and Hall conductance \cite{ref:streda}, $\sigma$, are given by
\begin{equation}\label{eq:rh-m-s}
\rho=\left(\frac{\partial \Omega}{\partial\mu}\right)_\varphi,\ m=-\frac 1{\Phi_0} \left(\frac{\partial \Omega}{\partial \varphi}\right)_\mu,\ \sigma =\frac {e^2}{h}\left(\frac{\partial \rho}{\partial \varphi}\right)_\mu .
\end{equation}
The magnetization and the Hall conductances of the two models are therefore related by:
\begin{eqnarray}\label{eq:m*-to-m}
m_l(\mu,T,\varphi^{-1})&=&\frac 1 {\Phi_0}\,\Omega_b(\mu,T,\varphi)-\varphi\,m_b(\mu,T,\varphi);\nonumber \\
\sigma_l(\mu,T,\varphi^{-1})&=&\frac {e^2}{h}\,\rho_b(\mu,T,\varphi)-\varphi\,\sigma_b(\mu,T,\varphi)
\end{eqnarray}
The duality relation of the Hall conductances is a generalization of a relation that follows from the Diophantine equation in~\cite{ref:tknn}.

The magnetization
is related to the Hall conductance by the Maxwell relation
\begin{equation}\label{eq:hall}
-\left(\frac{\partial m}{\partial\mu}\right)_\varphi=
\frac1\Phi_0\left(\frac{\partial\rho}{\partial \varphi}\right)_\mu =\frac 1 {ec}\,  \sigma .
\end{equation}
It follows that whenever the Hall conductance is quantized, the magnetization has a quantized slope that is an integer multiple of $1/\Phi_0$. In the Hofstadter model  $\sigma_b$ and $\sigma_l$ are quantized in the gaps \cite{ref:tknn,ref:bell}. It follows that the magnetization has quantized (and therefore also constant) slopes in the gaps.
  
We now turn to a more detailed study of the magnetization in the gaps at $T=0$, starting with $m_b$. For a fixed rational $\varphi$ this is a finite collections of lines. 
From Eqs.~(\ref{eq:symmetry-omega}),~(\ref{eq:rh-m-s}) it follows that $m_b$ is a symmetric function of~$\mu$.

When $\varphi=p/q$, with $p$ and $q$ relatively prime, the magnetization can be computed numerically. In this case the potential in Eq.~(\ref{eq:hofs-matrix}) is periodic with period $q$ and the spectral analysis of $H(\varphi,\xi)$ further reduces to the study of $q\times q$ matrices ${\Hm}(\varphi,\xi,\eta)$ labelled  by a second Bloch (quasi) momentum $\eta$. The magnetization $m_b(\mu,\varphi)$ in the gap of a split Bloch band can be shown to be given by 
\begin{equation}\label{eq:rational-magnetization}
-\frac q{2\pi\Phi_0}
\sum_{<,>}\int_0^{2\pi /q}\!\!{d\eta}\int_0^{2\pi/q}\!\!{d\xi}\,
{2\mu-(\varepsilon_>+\varepsilon_<)\over2\left(\varepsilon_>-\varepsilon_<\right)^2}
\,{\rm Im}
\big<u_>\big|\frac{\partial{\Hm}}{\partial \eta}
\big|u_<\big>\big<u_<\big|\frac{\partial{\Hm}}{\partial \xi}\big|u_>\big>
\ .\end{equation}
 $\varepsilon$  are the eigenvalues and  $u$ the corresponding eigenvectors of the matrix. $>,<$ refer to states above and below the gap. Our numerical results are based on this equation.

\begin{figure}[h]
\hskip0.5in \epsfxsize=4.0in \epsfbox{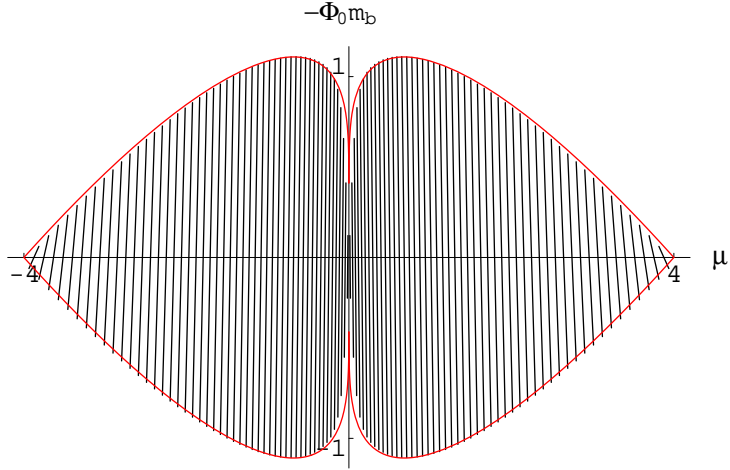}
\bigskip
\caption{The graph of the magnetization $m_b$ of a split Bloch band with $\varphi=1/101$. When $\varphi$ tends to zero the magnetization has a limiting envelope denoted by~$L_b(\mu,0)$.}
\label{fig:1/q-tb}
\end{figure}

{Our investigation of the magnetization in the gaps bears an interesting relation to the work of~\cite{ref:weigman}, who study of the energy $E$ per unit cell of the Hofstadter model. $E$ is the Legendre transform of $\Omega$,
\begin{equation}\label{eq:legendre} E(\rho,\varphi)=\mu\rho-\Omega(\mu,\varphi)\ ,
\end{equation}
and~\cite{ref:weigman} focused especially on its minima for fixed $\rho$. The properties of the Legendre transform imply that maxima of $\Omega$ as a function of $\varphi$, {\it i.e.}, points where $m=0$ and $\sigma>0$, are (local) minima of $E$. It follows from our semi-classical analysis (see below) that $E(\rho,\varphi)$ has a minimum in every gap for $\varphi=1/q$, see for example Fig.~\ref{fig:1/q-tb}. It also follows from Eq.~(\ref{eq:legendre}) that $E$ has a cusp as a function of $\varphi$ at the gaps, with tangent slopes given by the limiting values of $m$ at the ends of the gap.}

The Hofstadter model can be viewed as the result of Peierls  substitution~\cite{ref:spohn}, which stipulates that in a weak magnetic field the classical Hamiltonian 
\begin{equation}\label{eq:classical-hofstadter}
H_B(\xi,\eta)= 2\cos\xi+2\cos\eta
\end{equation}
is quantized by imposing $ [\xi,\eta]= 2\pi \,i\,\varphi$. Fluxes near $\varphi=0$ can then be analyzed by semiclassical methods where the role of $\hbar$ is played by  $2\pi\varphi$. The classical $\xi-\eta$ phase space is the two dimensional torus.   Since its area is $(2\pi)^2$, quantization of the torus is consistent provided the number of states, $1/\varphi$, is an integer. For this reason we shall assume that $\varphi=1/q$. 

The classical Hamiltonian $H_B(\xi,\eta)$ is a Morse function with one minimum, (at energy $-4$), one maximum, (at energy $4$), and two saddle points  (at energy zero). It follows that the  classical trajectories with $E\neq 0$ are contractible and bound a disc like domain. Only the level set  $E=0$ winds around the torus.

Let 
\begin{equation}\label{eq:action}
S(E)=\int_{H_B<E} d\xi\wedge d\eta
\end{equation}
 be the classical action associated with a trajectory of energy $E$.
By the Bohr-Sommerfeld quantization rule, the negative part of the spectrum is given, to leading order in $\varphi$, by those energies $E_n$ for which
\begin{equation}\label{eq:dispersion} 
S(E_n)=(n+\gamma)\varphi,\quad\gamma=1/2,\  \ n=1,\dots, \lfloor q/2\rfloor\end{equation}
This gives a good approximation to the spectrum provided that $\varphi$ is small and $E$ is  far from the separatrix. More precisely, one needs $\varphi\ll |E|$. The energy $E_n$ of Eq.~(\ref{eq:dispersion}) is then an approximation to the $n$-th energy band which is exponentially localized in energy near $E_n$. Near the separatrix one finds wider bands which we do not study~\cite{ref:mid-sc}. We therefore assume below that $\mu\ll -\varphi$. The magnetization for $\mu\gg \varphi$ then follows by Eq.~(\ref{eq:symmetry-omega}). The region which we do not study, $|\mu|=O(\varphi)$, becomes arbitrarily small in the limit $\varphi\to 0$. This is a limit we eventually take.

Fixing the chemical potential in the $n$-th gap with $\mu<0$, the semiclassical thermodynamic potential at $T=0$ is
\begin{equation} \label{eq:omega-1/q}
\Omega_b=\mu \rho-\sum_{j=0}^{n-1}E_j= \mu n\varphi -\varphi\,\sum_{j=0}^{n-1}\,S^{-1} \,\big( (j+\gamma)\varphi \big). \end{equation}
By the second Euler-Maclaurin formula, the sum can be approximated by an
integral (to $O(1/q^2)$),
\begin{equation}\label{eq:omega-semiclassic}
\Omega_b(\mu,\varphi)= \mu n
\varphi-\int_{0}^{\varphi n}S^{-1}(x)\,dx\, , \quad
\mu<0.\end{equation} 
The magnetization in the $n$-th gap is therefore
\begin{equation}\label{eq:M-sigma-B-semi}
m_b=-\frac1{\Phi_0}\,\big(\mu-S^{-1}(n \varphi)\big)n\ \end{equation}
and vanishes when $\mu=S^{-1}(n\varphi)$. Since $\gamma=1/2$ this happens in the middle of the gap, a property also evident in Fig.~\ref{fig:1/q-tb}.

Since the slope of $m_b$ in the $n$-th gap is $ n/ \Phi_0$, the magnetization
increases in this interval from $- n (E_n-E_{n-1})/2\Phi_0$
to $+ n (E_n-E_{n-1})/2\Phi_0.$ The extremal value $L_b$ of
the magnetization in the $n$-th gap is therefore
\begin{equation}L_b=\pm \frac {n\varphi} {2\Phi_0}\,{S^{-1}}'(n \varphi)\ .\end{equation}
Fixing the chemical potential $\mu$ and letting $\varphi\to0$, the
product $n\varphi$ approaches the limiting value $S(\mu)$,
see~(\ref{eq:dispersion}). The
envelope approaches
\begin{equation}\label{eq:envelop}L_b(\mu,\varphi=0)=
\pm{S(\mu)\over2\Phi_0S'(\mu)}\ .\end{equation} 
The graph of $L_b(\mu,0)$ is shown in Fig.~\ref{fig:1/q-tb} overlayed with numerical results for the magnetization that follow from the (exact) Eq.~(\ref{eq:rational-magnetization}). This graph is an analog of the de Haas-van Alphen oscillations in metals~\cite{ref:aschroft-mermin}. 

\begin{figure}[h]
\hskip0.6in \epsfxsize=4.0in \epsfbox{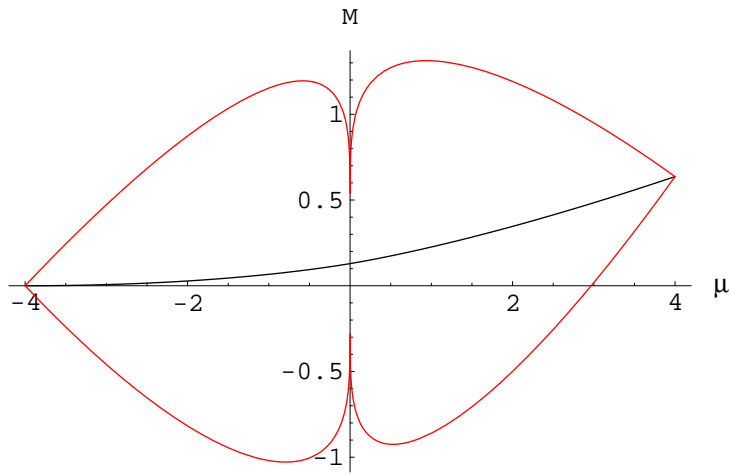} 
\bigskip \caption{ The central curve gives the limiting magnetization, $m_l(\mu,\varphi=\infty)$ and the upper and lower curves form the lip $L_l(\mu,\varphi=1)$}\label{fig:dualmag}
\end{figure}

$L_b(\mu,0)$ is the mother of all lips: We shall argue below that the lips at other (rational) fluxes are deformations of it, in both the the split-Bloch and split-Landau versions of the model. The lips associated with the split Landau level $L_l(\mu,1/n)$, with integer $n$, are particularly simple deformations of the mother of all lips. This can be seen as follows: From Eq.~(\ref{eq:m*-to-m}) 
\begin{equation}\label{eq:lips-dual}
  L_l(\mu,1/n)=\frac1\Phi_0\Omega_b(\mu,\infty)-nL_b(\mu,0)
\end{equation}
Since  $\Omega_b$ is a convex, and therefore continuous, function of $\mu$, Eq.~(\ref{eq:lips-dual}) describes a scaling of $L_b(\mu,0)$ and a shifting of its baseline, see Fig.~\ref{fig:dualmag}. The special case $m_l(\mu,\infty)$ ($n=0$) is different: The lip is scaled to zero width, and $m_l(\mu,\infty)$ is a continuous function of $\mu$. 

Although the actual shape of the lip $L_b(\mu,0)$ cannot be expressed in terms of elementary functions, its qualitative features can be readily understood and are determined by the critical points of the band function $H_B(\xi,\eta)$.  Let $\delta \mu$ denote the distance of $\mu$ from a band edge. Since $H_B$ is non-degenerate, the action, $ S(\mu)$ is  linear in $\delta\mu$ near the maximum and the minimum---as for the harmonic oscillator. This  says that the lip terminates linearly: 
\begin{equation}\label{eq:linpinch} S(\mu )\sim |\delta\mu|, \
L_b(\mu,0)\sim \pm \delta\mu\, .\end{equation}
Near the separatrix at $E=0$, which passes through the two saddle points of  $H(\xi,\eta)$, the density of states diverges logarithmically and $L_b$ vanishes logarithmically
\begin{equation}\label{eq:logpinch} 
S(\mu)-S(0)\sim \mu\log|\mu|,\ L_b(\mu,0)\sim 
\pm (\log\delta\mu)^{-1}.\end{equation}

\begin{figure}[h]
\hskip-1.in\epsfxsize=7.in \epsfbox{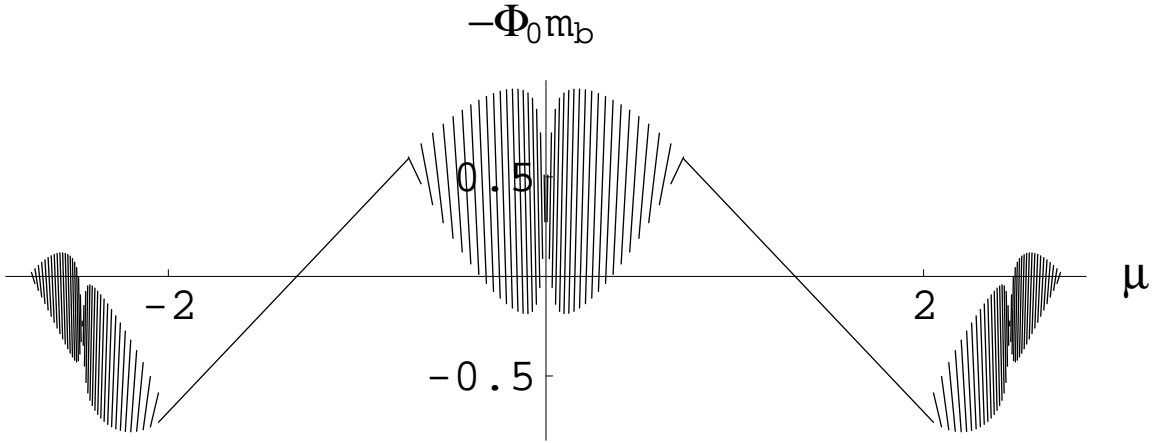} 
\caption{Plot of the magnetization as function of the chemical potential for flux that is  40/121 of the unit of quantum flux through the unit cell of the tight binding Hofstadter model. The magnetization is plotted for values of chemical potential that correspond to a spectral gap.} \label{fig:mg16/49}
\end{figure}
Consider now the magnetization for $\varphi$ close to a rational $p/q$. Since the gap edges are H\"older continuous of order 1/2 in $\varphi$ \cite{ref:mouches}, the $q-1$ gaps ($q-2$ for even $q$)  persist for nearby fluxes and so does the magnetization in these gaps. The $q$ bands, however, splinter.  This breakup can be described  by semiclassical methods as follows~\cite{ref:azbel-wilkinson}:
For rational flux $\varphi=p/q$,  the $q\times q$ matrix ${\Hm}(\varphi,\xi,\eta)$ gives rise to $q$ magnetic Bloch bands $E_j(\xi,\eta),\ j=1,\dots,q$. Each magnetic Bloch band  has one maximum, one minimum and two saddle points that have coinciding energies, and are linked by a separatrix. This is a consequence of Chambers relation \cite{ref:chambers}. For $\varphi$ close to $p/q$, the $q\times q$ matrix valued function is quantized by imposing the commutation relation $[\xi,\eta]=2\pi\,i\,(\varphi-p/q)$.

The semiclassical strategy leading to the qualitative description of the lip at $\varphi=0$ remains valid in this more general case, but the details are different. (For example, $\gamma$ does not take a universal value $1/2$ but rather becomes a function of $E$ and $\varphi$.) However, one still expects that the envelope of the magnetization is pinched linearly at the band edges and logarithmically at the separatrix. This gives all $L_b(\mu, p/q,j), \ j=1,\dots, q$, a lip-like shape, a distorted version of Fig.~\ref{fig:1/q-tb}. Getting explicit expressions for $L_b(\mu, p/q,j)$ is  a hard problem  which lies beyond the scope of this paper. However, numerical support for our claims is given in Fig.~\ref{fig:mg16/49}. Here the magnetization as a function of the chemical potential is shown for the flux $\varphi=\frac 1 {3+\frac 1 {40}}$ which is close to  $\frac 1 3$. The three bands of $\varphi=\frac 1 3$ split and give rise to rapidly oscillating magnetization with distorted lip-shaped envelopes.

{Band splitting occurs on all energy scales of the Hofstadter butterfly, because of its fractal nature. We have shown that band splitting is always accompanied by de Haas-van Alphen oscillations in the magnetization. The fractality of the spectrum is reflected by the fact that the magnetic oscillations occur on arbitrarily small energy scale. The fingerprints of fractality are most evident for flux ratios $\varphi$ which are irrational numbers that are well approximated by rationals (e.g. Liouville numbers). For such $\varphi$ the graph of the magnetization as a function of $\mu$, such as those shown in Figs.~\ref{fig:1/q-tb} and~\ref{fig:mg16/49}, is itself fractal, with magnetic oscillations bounded by lip-like envelopes appearing on infinitely many scales.}

{\bf Acknowledgment:} This work is supported by the Technion fund for promotion of research and by the EU grant HPRN-CT-2002-00277.




\end{document}